\documentclass{article}
\usepackage{spconf,amsmath,graphicx,bbding}
\usepackage{verbatim,amssymb,cite,booktabs}
\title{Multi-task Learning with Cross Attention for Keyword Spotting}
\name{Takuya Higuchi$^1$, Anmol Gupta$^2{}^{\ast}$, Chandra Dhir$^1$}
\address{$^1$Apple\\$^2$Department of Computer Science, The University of Hong Kong}
\begin{document}
%\ninept
%
\maketitle
\begin{abstract}
Keyword spotting (KWS) is an important technique for speech applications, which enables users to activate devices by speaking a keyword phrase.  Although a phoneme classifier can be used for KWS, exploiting a large amount of transcribed data for automatic speech recognition (ASR), there is a mismatch between the training criterion (phoneme recognition) and the target task (KWS).  Recently,  multi-task learning has been applied to KWS to exploit both ASR and KWS training data.  In this approach, an output of an acoustic model is split into two branches for the two tasks, one for phoneme transcription trained with the ASR data and one for keyword classification trained with the KWS data.  In this paper, we introduce a cross attention decoder in the multi-task learning framework.  Unlike the conventional multi-task learning approach with the simple split of the output layer,  the cross attention decoder summarizes information from a phonetic encoder by performing  cross attention between the encoder outputs and a trainable query sequence to predict a confidence score for the KWS task.  Experimental results on KWS tasks show that the proposed approach achieves a $12\%$ relative reduction in the false reject ratios compared to  the conventional multi-task learning with split branches and a bi-directional long short-team memory decoder.
\end{abstract}
\section{Introduction}
\label{sec:intro}
\renewcommand{\thefootnote}{\fnsymbol{footnote}}
\footnotetext[1]{Work performed at Apple}
\renewcommand*{\thefootnote}{\arabic{footnote}}

Keyword spotting (KWS) is a task to detect a keyword phrase from audio. KWS enables users to activate voice assistant systems on devices, such as smart phones and smart speakers, by simply speaking the keyword phrase. For  usability and privacy of users, it is important to deploy an accurate KWS system on-device.

A typical approach for KWS is to train a keyword-specific acoustic model to predict a confidence score for each keyword phrase. Earlier works use deep neural networks with a hidden Markov model (HMM) \cite{chen2014small, prabhavalkar2015automatic, panchapagesan2016multi,
sun2017compressed,kumatani2017direct,
guo2018time,wu2018monophone,sigtia2018}, 
and more recent works use convolutional neural networks (CNNs) \cite{sainath2015convolutional} and recurrent neural networks  \cite{fernandez2007application,arik2017convolutional}.
 Hardware friendly model architectures have also been investigated for small footprint KWS \cite{tucker2016model,alvarez2019end,higuchi2020stacked,yilmaz2020deep}. These models are typically trained on a KWS dataset, which consists of pairs of audio and corresponding phrase level labels. Although these approaches are a direct optimization for the target task, preparing a large labeled in-domain KWS dataset is challenging practically due to, e.g., the sparsity of false triggers and privacy concerns.
%  because KWS is used to decide if a voice assistant should be turned on and audio is sent to a server or not.

Another approach is to use an acoustic model of an automatic speech recognition (ASR) system. The ASR acoustic model (e.g., a phoneme classifier) is trained on a transcribed speech dataset (ASR dataset) to perform ASR \cite{zhuang2016unrestricted,rosenberg2017end,he2017streaming,adya2020hybrid}. At inference, a decoding score for a particular keyword phrase is computed, which corresponds to a confidence score of the presence of the keyword phrase. 
The advantage of this approach is that a large transcribed ASR dataset can be used for model training, and the keyword phrase is configurable at test time.

Recently, multi-task learning has been applied to KWS \cite{panchapagesan2016multi,9053577,sigtia2020progressive} to better generalize models leveraging both large ASR and in-domain KWS datasets. In this framework, an output layer of the acoustic model is split into two branches for the two tasks. Then the model is trained on both the ASR and KWS loss functions.

In this paper, we introduce a cross attention decoder in a multi-task learning framework. Unlike conventional multi-task learning, the cross attention decoder  summarizes information from the acoustic model (i.e., a phonetic encoder) using attention layers. Hidden representations from the phonetic encoder are fed into the cross attention decoder, and then cross attention between the phonetic representations and a query sequence is performed to predict a scalar confidence score for the KWS task. The phonetic encoder and the cross attention decoder are jointly trained using the multi-task learning framework. Experimental results on KWS tasks show that the proposed cross attention decoder outperforms the conventional multi-task learning and a bi-directional long short-term memory (BLSTM)-based decoder and achieves a $12\%$ relative reduction in the false reject ratios (FRRs) on average.

The remainder of this paper is organized as follows. Section \ref{sec:related} reviews related work and describes the contributions of this paper. Section \ref{sec:proposed} presents our proposed approach. Section \ref{sec:exp} describes experimental evaluation and Section \ref{sec:conc} concludes the paper.

\section{Related work}
\label{sec:related}
T. Bluche et al. also proposed to use a decoder on top of phoneme classifier outputs \cite{bluche2019predicting}. A BLSTM keyword encoder was trained to predict phrase level confidence scores based on outputs of an LSTM phoneme classifier and a specified keyword phrase.
% They obtain the confidence score for arbitrary keyword phrases by applying the convolutional filters on the outputs of the phoneme classifier. 
The LSTM phoneme classifier was pre-trained on an ASR dataset and fixed during training of the keyword encoder.  In contrast, our encoder and decoder are jointly trained from scratch, exploiting both ASR and KWS data in the multi-task learning framework. Moreover, we use an attention-based architecture for the decoder, following  recent successes of Transformers in ASR.
% The performance of the proposed cross attention decoder is compered with a BLSTM-based decoder in Section \ref{sec:exp}.

Transformers were originally proposed in \cite{vaswani2017attentionnips} and applied to ASR (e.g., \cite{karita2019comparative}). Adya et al. used the vanilla Transformer as a phoneme classifier for KWS \cite{adya2020hybrid}. Unlike the original Transformer decoder used in the previous articles, our cross attention decoder is not an auto-regressive model since our decoder predicts a scalar confidence score for a keyword phrase given an audio sequence. The length of a sequence of query vectors is fixed for our decoder, and the query vectors are jointly trained with the model parameters.

Tian et al. also applied multi-task learning for KWS, where they used both ASR and KWS data to train a recurrent neural network transducer model \cite{9414339}.  Only ASR data was used to train a prediction network so that the prediction network did not overfit to KWS data, and both ASR and KWS data were used to train an encoder.  Phoneme-level (or syllable-level) labels were used to train the encoder with both ASR and KWS data. In contrast, our proposed approach uses phoneme-level labels for ASR data, and phrase-level labels for KWS data. The proposed cross attention decoder summarizes phonetic information and performs a phrase-level prediction, which is used to compute a phrase-level loss on KWS data. The phrase-level prediction achieves better KWS performances compared to a phoneme-level prediction as shown in section \ref{sec:exp}.

\section{Cross attention for multi-task learning}
\label{sec:proposed}
\subsection{Overview}

\begin{figure}[t]
  \centering
  \includegraphics[width=\linewidth]{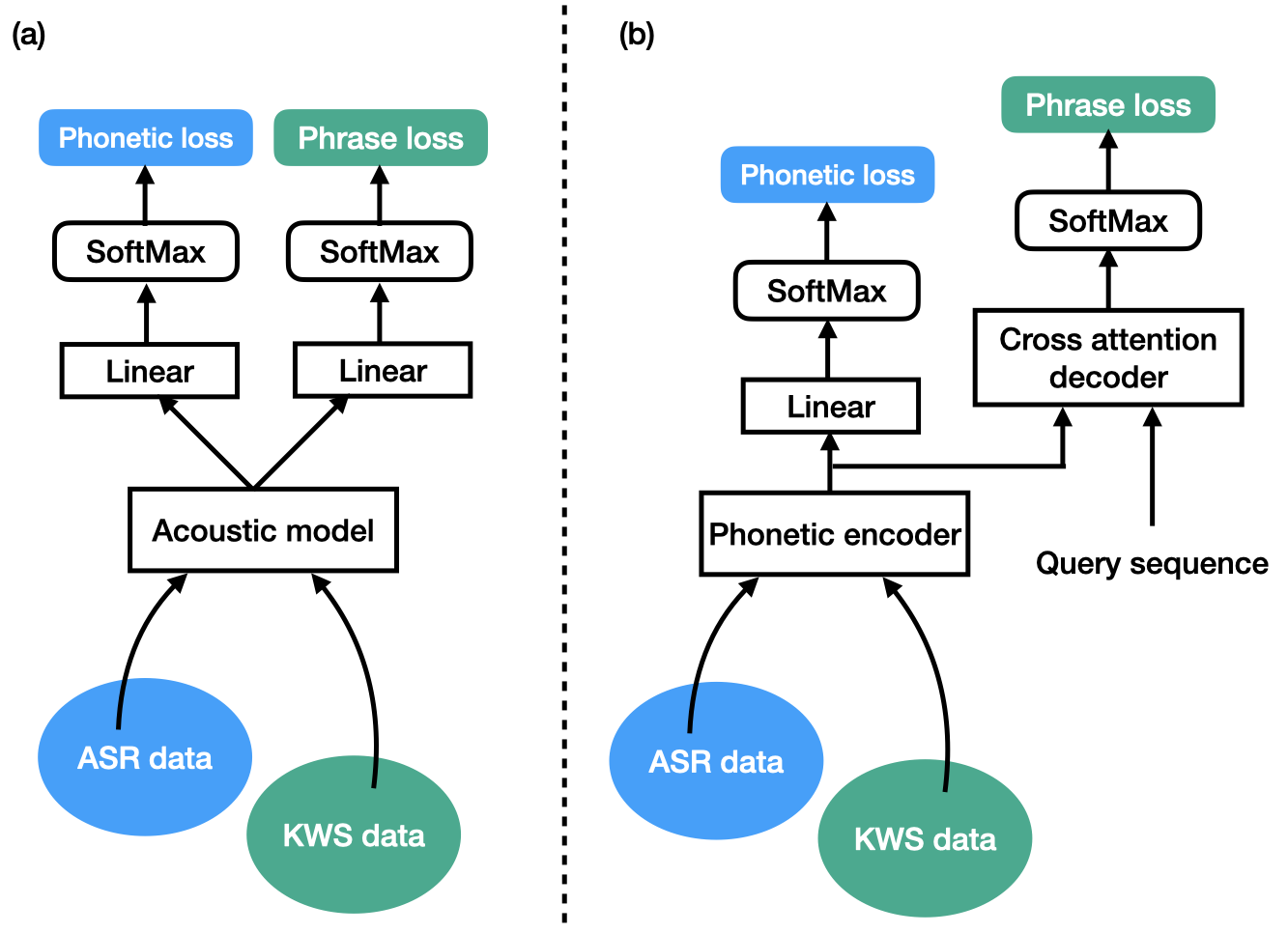}
  \caption{Block diagrams of (a) conventional multi-task learning for KWS \cite{9053577,sigtia2020progressive} and (b) our proposed approach.  In the conventional approach , a last layer is simply split into two branches, one for  phoneme prediction and one for phrase prediction. In contrast,  the proposed approach uses a phonetic encoder for phoneme prediction,  and a cross attention decoder is introduced to efficiently summarize phonetic information for phrase prediction.}
  \label{fig:overview}
\end{figure}

Figure \ref{fig:overview} shows an overview of the conventional multi-task learning approach and the proposed approach.  In the conventional approach, the output of the acoustic model is split and fed into two different branches, one for phoneme classification and one for KWS. Then the model is trained on both ASR data and KWS data using either phonetic loss or discriminative loss, depending on which dataset a data sample is from.  Unlike the multi-branch approach in the conventional multi-task learning, we introduce the cross attention decoder which works on top of the phonetic encoder outputs.  The decoder takes the outputs from the phoneme classifier as key and value vectors for attention layers.  Then cross attention is performed between the phonetic encoder outputs and a trainable query sequence to predict a score for KWS.  The encoder and the decoder are jointly trained in the multi-task learning framework.

\subsection{Multi-task learning}

In the multi-task learning framework, the model is trained using both phonetic loss and phrase loss \cite{panchapagesan2016multi,9053577,sigtia2020progressive}.  Let us assume that we sample $N$ utterances for a mini-batch from a combined set of an ASR dataset and a KWS dataset.  $J$ utterances are sampled from the KWS dataset and $(N-J)$ utterances from the ASR dataset. The objective function on a mini-batch for training can be written as
\begin{equation}
\mathcal{L} = \mathcal{L}^{(phone)} + \alpha \mathcal{L}^{(phrase)}, \label{eq:MTL}
\end{equation}
where $\mathcal{L}^{(phone)} = \frac{1}{(N-J)}\sum_{n=1}^{(N-J)} \mathcal{L}_{n}^{(phone)}$ and $\mathcal{L}^{(phrase)} = \frac{1}{J}\sum_{j=1}^{K} \mathcal{L}_{j}^{(phrase)}$ denote phonetic and phrase losses computed on utterances from the ASR dataset and the KWS dataset, respectively. $\alpha$ is a scaling factor for balancing the phonetic loss and the phrase loss. To train the acoustic model on these two losses, an output layer of an acoustic model is typically split into two branches, one for the phonetic loss and one for the phrase loss.

\subsection{Cross attention decoder}
\begin{figure}[t]
  \centering
  \includegraphics[scale=0.25]{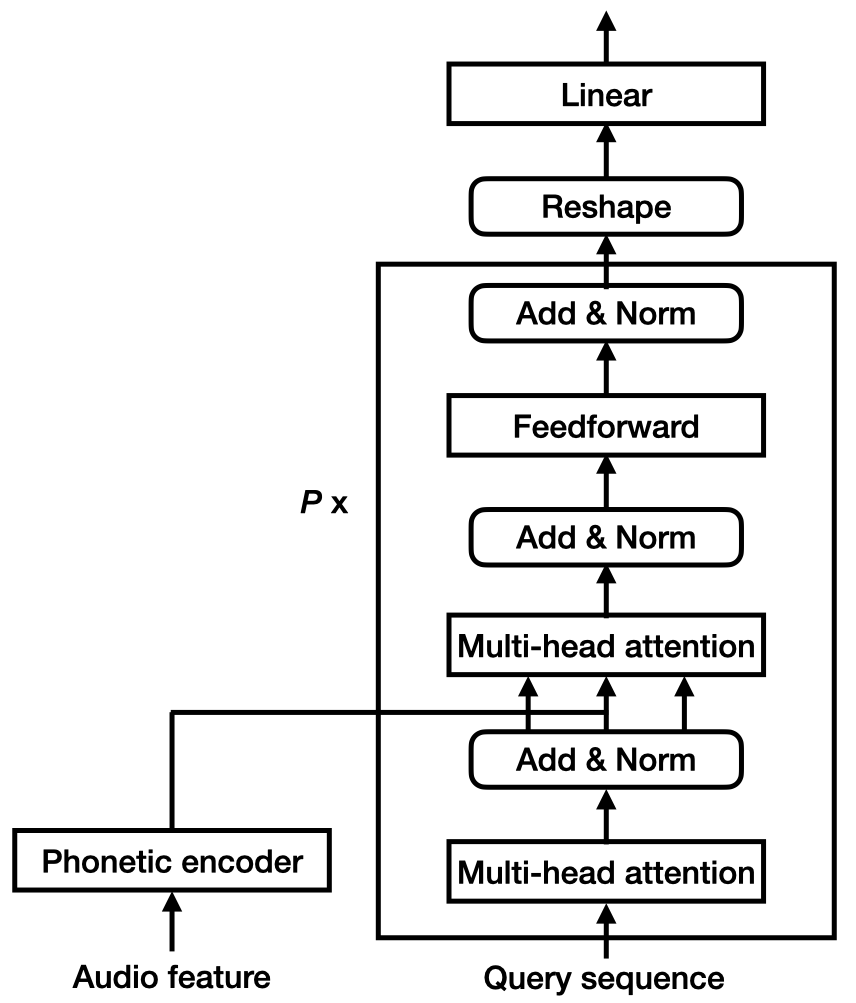}
  \caption{The proposed cross attention decoder.}
  \label{fig:decoder}
\end{figure}

We introduce a cross attention decoder to perform phrase level classification based on phonetic predictions from the acoustic model, i.e., the phonetic encoder.
% First, the encoder is trained with the CTC loss on phonetic labels.
  Let $\mathcal{X}_{n} = \lbrace x_{n,t} | t = 1, ..., T\rbrace$ denote a set of input features at time frames $1$ to $T$ for $n$-th utterance sampled from the ASR dataset.  The phonetic encoder transforms the input feature sequence into a sequence of hidden representations as
\begin{equation}
\mathcal{H}_{n} = Encoder(\mathcal{X}_{n}),
\end{equation}
where $\mathcal{H}_{n} = \lbrace h_{n,t} | t = 1, ..., T\rbrace$ denotes a set of the hidden representations, and $Encoder(\cdot)$ denotes a mapping function by the phonetic encoder defined with neural networks.
% We use self-attention-based acoustic model \cite{} for the encoder.
  Then a linear layer is applied to project the hidden representations to logits.
%\begin{equation}
%\mathcal{\hat{Y}}_{n} = Linear(\mathcal{H}_{n}),
%\end{equation}
%where $\mathcal{\hat{Y}}_{n} = \lbrace \hat{y}_{n,t} | t = 1, ..., T\rbrace$ denotes a set of logits for phonetic classes.
  Next the CTC loss for the $n$-th utterance $\mathcal{L}_{n}^{(phone)}$ is computed using the logits and phoneme labels.

In addition, a discriminative phrase level loss is computed using the cross attention decoder on utterances sampled from the KWS dataset.  
 Let us assume that $j$-th utterance is sampled from the KWS dataset, and $\mathcal{X}_{j} = \lbrace x_{j,t} | t = 1, ..., T\rbrace$ denotes acoustic features of the $j$-th utterance. The features are first processed by the phonetic encoder as
\begin{equation}
\mathcal{H}_{j} = Encoder(\mathcal{X}_{j}),
\end{equation}
to produce a sequence of the hidden representations, $\mathcal{H}_{j}$.

Following the recent success of Transformers in ASR,  our cross attention decoder is based on Transformer blocks with attention layers. The attention layer for a query matrix $Q$, a key matrix $K$ and a value matrix $V$ can be written as
\begin{equation}
Attention(Q,K,V) = softmax(\frac{QK^{T}}{\sqrt{d}})V,
\end{equation}
where $d$ is the size of the key and the query vectors. 
 Figure \ref{fig:decoder} shows a detailed block diagram of the cross attention decoder. The decoder takes the encoder output $\mathcal{H}_{j}$ and a sequence of trainable vectors (i.e., query vectors) as inputs. Let $\mathcal{Q} = \lbrace q_{m} | m = 1, ..., M\rbrace$ denote a sequence of the query vectors, where $q_{m} \in \mathbb{R}^{d \times 1}$. First, self-attention is performed on $\mathcal{Q}$ by a multi-head attention layer as
\begin{equation}
 \mathcal{H}^{q} = MHA(Q,K,V), 
 \end{equation}
 where $\mathcal{H}^{q}$ denotes the output from the self-attention layer and $MHA(Q,K,V)$ denotes the multi-head attention layer described as
\begin{equation}
 MHA(Q,K,V) = Concat(head_{1},...,head_{h}), 
 \end{equation}
where $head_{h} = Attention(Q,K,V)$. $Q$, $K$, and $V$ are obtained by applying affine transformations to $\mathcal{Q}$, and different heads use different affine transformations. Then, cross attention is performed with the output of the self-attention layer and the phonetic encoder output $\mathcal{H}_{j}$ by $MHA(Q',K',V')$, where $Q'$ is obtained by applying an affine transformation to $\mathcal{H}^{q}$, and $K'$ and $V'$ are obtained by applying affine transformations to $\mathcal{H}_{j}$.  The output of the cross attention layer is fed into position-wise feedforward networks. Residual connection and layer normalization \cite{ba2016layer} are used after each attention block and feedforward block following the original Transformer.  The transformer block is repeated $P$ times.   Finally, the output from the Transformer block is reshaped and fed into a linear layer to predict a scalar logit for the keyword phrase. Unlike the original Transformer architecture, our decoder for KWS is not an auto-regressive model and the length of the query sequence, $M$, is fixed, which enables reshaping so a linear layer can produce a scalar logit for each keyword phrase per audio sequence.  Moreover,  positional encoding is not required for the query vectors since the query vectors are jointly optimized with model parameters and then fixed for any audio input at inference.
  
The phrase loss for the $j$-th utterance, $\mathcal{L}_{j}^{(phrase)}$, is defined as the cross entropy between the logits from the decoder and utterance-wise phrase labels. The encoder and the decoder are jointly trained in the multi-task learning framework using Eq. (\ref{eq:MTL}).

\section{Experimental evaluation}
\label{sec:exp}

We evaluated the effectiveness of the proposed approach on a KWS task, and compared its performance with a self-attention phonetic decoder with/without the conventional multi-task learning and a BLSTM decoder with the conventional multi-task learning. Although we used our internal datasets in experiments,  our proposed approach is easily applicable to any public ASR and KWS datasets.

\subsection{Data}
Our ASR training data consisted of approximately 3 million utterances of transcribed near-field speech signals recorded with devices such as smart phones. Then data augmentation was performed by convolving room impulse responses (RIRs) with speech signals. The RIRs were recorded in various rooms with smart speakers with six microphones. Additionally, echo residuals were added to the augmented data. As a result, we obtained approximately 9 million augmented utterances consisting of the near-field signals, simulated far-field signals, and simulated far-field signals with the echo residuals. The KWS data consisted of approximately $65k$ false triggers and $300k$ true triggers spoken by anonymous speakers, which were triggered by a reference voice triggering system. The audio signals were recorded with smart speakers. The KWS data were combined with the augmented ASR dataset, and utterances were randomly sampled from the combined dataset for mini-batch training.

For evaluation, we used two different datasets. The first  is a \emph{structured} dataset, where positive samples containing a keyword phrase were internally collected in controlled conditions from 100 participants, approximately evenly divided between males and females. Each subject spoke the keyword phrase followed by prompted voice commands to a smart speaker. The recordings were made in four different acoustic conditions: quiet, external noise from TV or kitchen appliances, music playing from the device at medium volume, and music playing at loud volume. 13000 such positive utterances were collected. For negative data, we used a set of 2000 hours of audio recordings which did not contain the keyword phrase by playing podcasts, audiobooks, TV play-back, etc. These negative audio samples were also recorded by the same smart speaker. The negative audio data allowed us to compute false accept (FA) per hour. The second dataset, called take home evaluation set, is a more realistic and challenging  dataset collected at home by employees. Each of the 80 participants volunteered to use the smart speaker daily for two weeks. By applying extra audio logging on device and personal review by the user, audio below the usual on-device trigger threshold was collected. This setup allowed us to collect challenging negative data, which was similar to the keyword phrase. We collected 7896 positive and 20919 negative audio samples\footnote{The amount of the dataset has been increased by additional participants compared to the evaluation dataset used in \cite{adya2020hybrid} and \cite{9053577}, so the result reported in this paper is not directly comparable.} for evaluation. This dataset allowed us to compute the absolute number of false accepts (FAs).

\subsection{Two-stage approach for efficient KWS}
\begin{figure}[t]
  \centering
  %\vspace{-8pt}
%\setlength{\abovecaptionskip}{5pt}
  \includegraphics[scale=0.45]{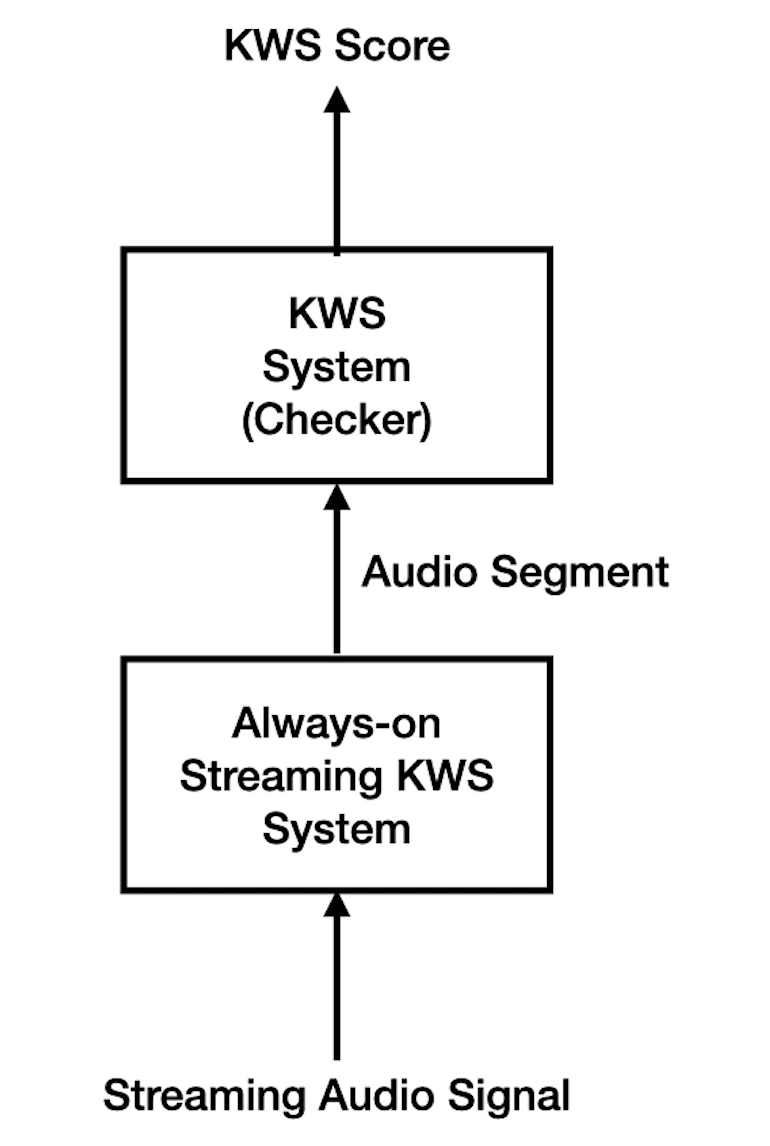}
  \caption{A two-stage approach for efficient KWS \cite{gruenstein2017cascade,sigtia2018}. A 1st pass light-weight KWS system is always-on and takes streaming audio signals, where a DNN-HMM system is used to obtain a KWS score and an alignment for an audio segment containing a keyword. Once the 1st pass KWS score exceeds a threshold,  the audio segment is passed to a bigger KWS model (so-called checker) and a KWS score is re-computed.}
  %\vspace{-8pt}
  \label{fig:TwoPass}
\end{figure}

We used a two-stage approach for efficient KWS \cite{gruenstein2017cascade,sigtia2018} as shown in figure \ref{fig:TwoPass}. A light-weight model was always-on and first detected candidate audio segments from streaming audio inputs. Once the segments were detected, a bigger model (so-called checker) was turned on and checked if the segments actually contained the keyword phrase or not. This two-stage approach greatly reduces compute cost and battery consumption on-device. For the 1st pass model, we used five layers of fully-connected neural networks with 64 hidden units as the acoustic model.  We used 20 target classes for the acoustic model; 18 phoneme classes for the keyword, one for silence and one for other speech.  We computed a 13-dimensional MFCC feature at a rate of 100 frames per second,  and supplied 19 consecutive frames to the acoustic model.  The confidence scores for KWS and alignments to extract audio segments were obtained using an HMM.  Given keyword start and end times from the HMM alignment,  we used $(start~time - 0.5)$ seconds  and $(end~time + 0.3)$ seconds for segmentation to ensure that the segment contained the detected keyword portion. The 1st pass threshold was set to obtain approximately 21 FA/hr on the structured evaluation dataset. We used the same 1st pass system for all the experiments and evaluated the effectiveness of our proposed model as the checker.

\subsection{Model training}
For a baseline phoneme classifier, we used a self-attention based acoustic model. The model consisted of 6 layers of Transformer blocks, each of which had a multi-head self-attention layer with 256 hidden dimension and 4 heads, followed by a feedforward neural network with 1024 hidden units. Finally, outputs from the Transformer blocks were projected to 54-dimensional logits for phonetic and blank labels by a linear layer. The baseline model was trained with the CTC loss\footnote{In \cite{adya2020hybrid}, the vanilla Transformer decoder was also trained along with the self-attention encoder using cross entropy loss, and used as a regularizer during training. We omitted the regularization just because of simplicity in our experiments. The regularization can be applied to all the approaches in our experiments including the proposed approach.}. The same architecture was also used for the conventional multi-task learning \cite{9053577} by splitting the last layer into 54 outputs for the phonetic CTC loss and three discriminative outputs for a positive class, a negative class and a blank label for the phrase level CTC loss. Regarding the proposed approach, we used the same self-attention phoneme classifier for the phonetic encoder. The cross attention decoder consisted of a Transformer decoder block (i.e., $P=1$) which had the same configuration as the Transformer blocks of the encoder except the cross attention block. The dimension of the query vector and the length of the query sequence were set at 256 and 4, respectively. The last linear layer projected the reshaped $1024 (256\times4)$-dimensional vector to two logits for positive and negative classes. The encoder and the decoder were jointly trained using the phonetic CTC loss and the phrase level cross entropy loss (see Section \ref{sec:proposed}). We also explored a BLSTM decoder by replacing the cross attention decoder by a layer of BLSTMs with 256 hidden units followed by a linear layer which processed a concatenated BLSTM outputs at the first and last frame to predict logits. The scaling factor $\alpha$ in Eq. (\ref{eq:MTL}) for the multi-task learning was experimentally set at $10$. $40$-dimensional log mel-filter bank features $\pm$ 3 context frames were used as inputs. In addition, we sub-sampled the features once per three frames to reduce computational complexity.
%\subsection{Training hyper-parameters}
 All models were trained using the Adam optimizer \cite{kingma2014adam}. The learning rate was first increased linearly to $0.0008$ until epoch $2$, then linearly decayed to $0.00056$ until epoch $16$. Finally the learning rate was exponentially decreased until the last epoch which was set at $28$. We used 16 GPUs for training and the batch size was 128 at each GPU.

\subsection{Results}

\begin{figure}[t]
  \centering
  %\vspace{-8pt}
%\setlength{\abovecaptionskip}{5pt}
  \includegraphics[width=\linewidth]{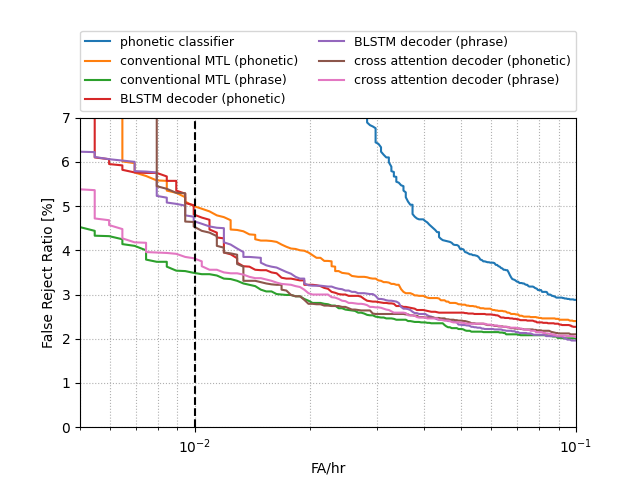}
  \caption{DET curves for structured evaluation set. The vertical dotted line indicates an operating point.}
  %\vspace{-8pt}
  \label{fig:det_edc}
\end{figure}

\begin{figure}[t]
  \centering
  %\vspace{-8pt}
%\setlength{\abovecaptionskip}{5pt}
  \includegraphics[width=\linewidth]{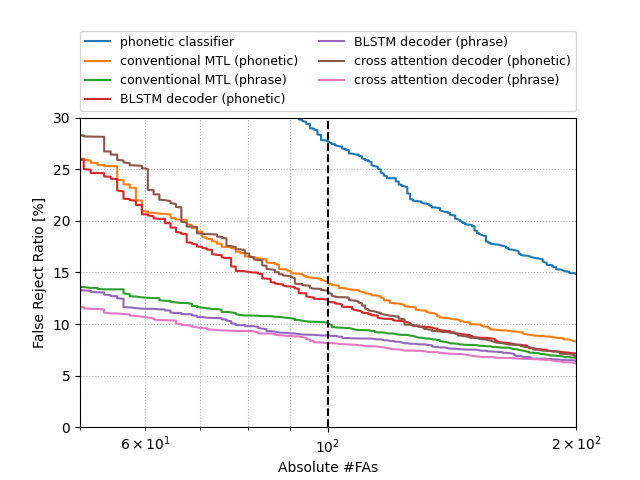}
  \caption{DET curves for take home evaluation set. The vertical dotted line indicates an operating point.}
  %\vspace{-8pt}
  \label{fig:det_thk}
\end{figure}

\begin{table*}[t]
  \caption{False reject ratios for structured evaluation set [$\%$] at an operating point of 1 FA/100 hrs, and for take home evaluation set at an operating point of 100 FAs.}
  %\vspace{-3mm}
  \label{tab:FRRs}
   \centering
   %\scalebox{0.8}{
\begin{tabular}{cccccc}
  \toprule
 &  MTL & Branch & Structured evaluation set&  Take home evaluation set & Avg.\\
  \midrule
 Phoneme classifier &  &Phonetic& 20.26 & 27.72 & 23.99\\ \midrule
 Conventional MTL \cite{9053577}& \checkmark &\begin{tabular}[c]{@{}c@{}}Phonetic\\ Phrase\end{tabular}& \begin{tabular}[c]{@{}c@{}}5.00 \\\textbf{3.49}\end{tabular}& \begin{tabular}[c]{@{}c@{}}14.11 \\10.11\end{tabular} &\begin{tabular}[c]{@{}c@{}}9.56 \\6.80\end{tabular}\\ \midrule
 BLSTM decoder & \checkmark &\begin{tabular}[c]{@{}c@{}}Phonetic\\ Phrase\end{tabular}& \begin{tabular}[c]{@{}c@{}}5.02 \\4.76\end{tabular} & \begin{tabular}[c]{@{}c@{}}12.36 \\8.89\end{tabular} &\begin{tabular}[c]{@{}c@{}} 8.69\\6.83\end{tabular}\\ \midrule
 Cross attention decoder & \checkmark &\begin{tabular}[c]{@{}c@{}}Phonetic\\ Phrase\end{tabular}& \begin{tabular}[c]{@{}c@{}}4.64 \\3.82\end{tabular} & \begin{tabular}[c]{@{}c@{}}13.21 \\\textbf{8.17}\end{tabular} &\begin{tabular}[c]{@{}c@{}} 8.93\\\textbf{6.00}\end{tabular}\\
     \bottomrule
\end{tabular}
%}
\end{table*}

Figures \ref{fig:det_edc} and \ref{fig:det_thk} show detection error tradeoff (DET) curves for all models evaluated on the structured evaluation dataset and take home evaluation dataset, respectively. The horizontal axis represents FA/hr for the structured dataset or the  absolute number of FAs for take home dataset. The vertical axis represents FRRs. Table \ref{tab:FRRs} shows FRRs obtained with the baseline and proposed models at operating points. In the case of multi-task learning, results from both the phonetic and  phrase branches were reported. First, multi-task learning significantly improved the FRRs compared to the phoneme classifier which was trained only on the ASR data. This result shows the effectiveness of using both the ASR and the KWS data for KWS model training. Second, the phrase branch always yielded better results than the phonetic branch, presumably because the phrase branch was directly optimized for the target task. Note that although the performance of the phonetic branch was not as good as the phrase branch, the phonetic branch has an advantage of flexibility where the keyword phrase is configurable at test time.
% Thirdly, the phrase branch of the BLSTM decoder and the cross attention decoder outperformed the conventional multi-task learning, presumably because the decoders were able to efficiently summarize information from the phoneme classifier in order to predict phrase level confidence scores.
 Lastly, the proposed cross attention decoder with the phrase branch yielded the best performance and achieved a $12\%$ relative reduction in the FRRs  compared to the conventional multi-task learning and the BLSTM decoder. The cross attention decoder has another advantage over the BLSTM decoder, which is less training time and less runtime cost as reported in \cite{adya2020hybrid}.

Even though the proposed decoder can effectively learn from the KWS training data\footnote{Cross validation loss with the conventional multi-task learning was $1.5\times$ higher than the loss with the cross attention decoder.}, the proposed approach with the phrase branch did not outperform the conventional multi-task learning for the structured evaluation set. This performance degradation could be because of mismatched conditions/distributions between the KWS training data and the structured evaluation dataset that was recorded in the controlled conditions.

\section{Conclusions}
\label{sec:conc}
We proposed the cross attention decoder in the multi-task learning framework for KWS. The cross attention decoder performed cross attention between the hidden representations from the phonetic encoder and the query sequence, and then predicted a confidence score for the KWS task. The phonetic encoder and the cross attention decoder were jointly trained in the multi-task learning framework leveraging both the ASR and KWS datasets. The proposed approach achieved a $12\%$ relative reduction in the FRRs compared to the conventional multi-task learning and the BLSTM decoder. Our future work includes an extension of this approach to open vocabulary KWS.
% where the keyword phrase is configurable at test time.

 \bibliographystyle{IEEEbib}
\bibliography{mybib}

\end{document}